\def\g{\gamma}
\def\r{\rho}
\def\f{\phi}
\def\H{{\dot a\over a}}
\def\hq{{{\dot a}^2\over a^2}}
\def\ra{\r(a)}

\def\gt{\hat\g}
\def\gy{\gt(y)}
\def\gyp{\gt(y')}
\def\ga{\g(a)}
\def\gap{\g(a')}
\def\gr{\g(\r)}

\def\der#1#2{{d#1\over d#2}}

\def\mez{{1\over2}}
\def\ter{{1\over3}}
\def\df{{\dot\f}}
\def\potf{V(\f)}
\def\pota{V(a)}

\def\su{\over}

\def\pg{{8\pi G\over3}}

\def\gp{{3\over8\pi G}}

\def\ln{\log(a/a_{i})}

\newcommand{\bee} {\begin{equation}}
\newcommand{\ene} {\end{equation}}
\newcommand{\beqa} {\begin{eqnarray}}
\newcommand{\enqa} {\end{eqnarray}}
\newcommand{\beqsa} {\begin{eqnarray*}}
\newcommand{\enqsa} {\end{eqnarray*}}
\newcommand{\bea} {\begin{array}}
\newcommand{\ena} {\end{array}}

\documentstyle[12pt]{article}
\begin{document}
\baselineskip=18pt

{{\hfill Preprint DSF-T-26/94}}\\

\begin{center}
{\Large\bf Inflaton Potential Reconstruction and
Generalized Equations of State}\\
\end{center}

\bigskip\bigskip

\begin{center}
{\bf G. Mangano, G. Miele} and {\bf C. Stornaiolo}

\end{center}

\bigskip

\small
\noindent
{\it Dipartimento di Scienze Fisiche, Universit\`a
di Napoli - Federico II -, Mostra D'Oltremare Pad. 19, 80125, Napoli, Italy,
and INFN, Sezione di Napoli, Mostra D'Oltremare Pad. 20, 80125, Napoli, Italy.}

\bigskip\bigskip\bigskip

\begin{abstract}
We extend a previous analysis concerning cosmological fluids with
generalized equations of state in order to study
inflationary scenarios. In the framework of the slow-roll
approximation we find the expressions for the perturbation parameters
$\epsilon$, $\eta$ and the density perturbation spectra in terms of
the adiabatic index $\ga$ as a function of the universe scale factor.
This connection allows to find straightforwardly $\ga$ corresponding,
for example, to the simplest {\it chaotic} model and to the
Harrison-Zeldovich potential and shows its capability to be applied to
more complicate situations. Finally, we use this description to
develop a new approach to the early universe dynamics, based on
a $1/N$ expansion, where $N$ is the e-fold number.
To this aim, we introduce a set of suitable
dimensionless variables and show that at the zero-th order in $1/N$, an
improved slow-roll approximation is obtained.
\end{abstract}

\noindent

\normalsize
\newpage

\section{Introduction}

The idea of deriving the inflaton potential by imposing particular
theoretical or phenomenological requirements has been widely discussed
in the recent literature. In \cite{1a}, the potential $\potf$ is deduced
by explicitly assuming the time evolution of the universe scale factor
$a=a(t)$,
and substituting it into the Friedmann-Robertson-Walker equations. The
resulting
potential is given in a parametric form, using the time as the parameter. In
\cite{2a} the hamiltonian approach was followed to postulate that the scalar
field hamiltonian is some function of $\log(a)$, $a$ being the
average scale factor in Bianchi models I and V. Finally in
\cite{3a} it is shown that, for any arbitrary equation of state, $\potf$ can
be related to the solution of a Riccati equation. More recently other methods
have been proposed. In particular in \cite{3b} it is shown how a family
of inflaton potentials can be reconstructed from the scalar density fluctuation
spectrum. This approach is improved in \cite{4a} where it is shown how the
potential can be uniquely determined from the knowledge of the tensor
gravitational wave spectrum. These spectra may be obtained  from the
detection of fluctuations in the temperature distribution in the Cosmic
Microwave Background Radiation (CMBR). These results, which are not general
as they are obtained in the slow-roll approximation, are corrected up to
second-order in slow roll parameters \cite{5a}.

In this letter the scalar field potential reconstruction
procedure contained in \cite{6a}, is applied to inflationary scenarios.
In particular, the properties of some relevant models are
related to a suitable class of equations of state for the cosmological fluid.
In section 2 the potential reconstruction procedure proposed in \cite{6a} is
briefly reviewed. This formalism allows to relate in a very simple way the
adiabatic index for the cosmological fluid considered to the slow-roll
parameters $\epsilon$ and $\eta$ \cite{4a}.
In section 3, we obtain the expression of fluctuation amplitudes
and spectra in terms of the adiabatic index ($\ga$), and for some relevant
inflationary potentials, i.e. the most simple chaotic theory $V(\phi)=\mu^2
\phi^2$, and the Harrison Zeldovich model we determine $\ga$.
Thus, in section 4, by using the approach outlined in section 2 we develop the
$1/N$ expansion for some relevant quantities concerning the early
universe dynamics, where $N$ is the e-fold number of the inflationary
stage. Remarkably, at the zero-th order in this parameter we get an
improved slow-roll approximation which contains as a particular case
the usual slow-roll regime. Finally in section 5 we give our conclusions.

\section{Potential reconstruction from the equation of state}

In a homogeneous and isotropic universe, whose metric can be cast
in the standard form
\bee
ds^2= dt^2-a^2(t) \left\{{dr^2 \over 1-kr^2}  +r^2 d\theta^2 + r^2
\sin^2{\theta} d\varphi^2 \right\}~~~,
\label{a-2}
\ene
the corresponding Friedmann-Robertson-Walker equations are
\bee
\hq+{k\over a^2}={8\pi G\over3}~\r~~~,
\label{a-1}
\ene
and
\bee
2{\ddot a\over a}+\hq+{k\over a^2}=-8\pi G~ p~~~,
\label{a0}
\ene
where $k$ can be chosen to be $+1$,
$-1$, or $0$ for spaces of constant positive, negative, or zero spatial
curvature respectively\footnote{in units $\hbar=
c=1$, $G=1/m_{Pl}^2$.}.\\
For a perfect fluid corresponding to a minimally
coupled self-interacting scalar field the energy density $\r$ and the
pressure $p$ read
\bee
\r=\mez\df^2+V(\f)~~~,
\label{a1}
\ene
\bee
p=\mez\df^2-V(\f)~~~,
\label{a2}
\ene
where $V(\f)$ is the potential of the self-interacting scalar field. By using
(\ref{a1}) and (\ref{a2}) the Klein-Gordon equation
\bee
\ddot\f+3\H\df+ \frac{ \delta V( \phi) }{\delta \phi} =0~~~,
\label{a3}
\ene
can be derived from the  conservation law
\bee
\dot \r=-3\H\left(\r+p\right)~~~.
\label{a4}
\ene
Hence from  $(\ref{a4})$, dividing by $\dot a$,
we can obtain the equation of state $p=p(\r)$
for the fluid. In general due to the arbitrariness of the
potential, we expect this equation to be non linear
\bee
p=\left(-\ter{a\su \r}\frac{d\r}{da}-1\right)\r\equiv(\gr-1)\r~~~.
\label{a5}
\ene
We note that, by virtue of $(\ref{a4})$, to give $\gr$ is equivalent to
assign $\ga$, by which we can describe the most general family of
inflationary potentials. This procedure, in fact, allows us to find a
parametric form for the scalar potential in terms of
the equation of state in a very simple and intuitive way, where
the expansion factor $a$ is the parameter.

{}From the definition of $\g$ contained in (\ref{a5}),
we obtain
\bee
\ra=\r_0\exp\left\{-3\int_{a_i}^a{\gap \over a'}da'\right\}~~~,
\label{a6}
\ene
where $a_{i}$ and $\rho_{0}=\rho(a_{i})$ are the initial values for
the scale factor and the energy density respectively. Eqs.
(\ref{a4}) and (\ref{a5}) together with (\ref{a6}) give,
\bee
\mez\df^2-\potf=(\ga-1)\left(\mez\df^2+\potf\right)~~~.
\label{a7}
\ene
Thus, the expressions for the kinetic energy and for the
potential follow
\bee
\mez\df^2={\ga\over2}\ra~~~,
\label{a8}
\ene
\bee
\pota={2-\ga\over2}~\ra={2-\ga\over2}~
\r_0~\exp\left\{-3\int_{a_i}^a{\gap\over a'}da'\right\}~~~.
\label{a9}
\ene
In Eq. (\ref{a8}), the factor $\ga/2$ can be seen as the weight of the
kinetic term with respect to the total energy density. By virtue of
(\ref{a8}) $\ga$ is bound to be positive, and it is exactly equal to
zero for pure exponential inflation.

To obtain the evolution of $\f$ with respect to $a$, let us
consider the square root of (\ref{a8})
\bee
\df=\pm\sqrt{\ga\ra}~~~.
\label{a10}
\ene
By comparing (\ref{a10}) with
\bee
\df=\der\f a\dot a= a\der\f a\sqrt{\pg\ra}~~~
\label{a11}
\ene
(we are assuming a spatially flat expanding universe), we get
\bee
\f(a)=\f(a_i)\pm \sqrt{\gp} \int^a_{a_i}{\sqrt{\gap}\su a'}da'~~~.
\label{a12}
\ene
Note that both (\ref{a9}) and (\ref{a12}) are completely general, and determine
$\potf$ once we assign $\rho_{0}$, and the function $\ga$. Thus the above
description provides a natural classification scheme for single-fluid
inflationary models, given in terms of the $\ga$ function. Note moreover that
by construction (\ref{a9}) and (\ref{a12}) formally solve (\ref{a3}), and that
the inflationary phase is characterized by $\ddot a>0$ \cite{7a}, \cite{8a},
and therefore, whenever $\ga<2/3$.

The last step in reconstructing the potential is the determination of the
parameter $\r_0$. We can see that this problem is related to the {\it
efficiency} (how fast the inflation occurs)
of the inflationary process devised here. The function $\ga$ contains the
information of the amount of growth of the universe during inflation,
and also allows to obtain explicitly the inflationary time interval
$\Delta t$ in terms of $\r_0$
\bee
\Delta t= \sqrt{3 \over 8\pi G\r_0}~
\int^{a_{f}}_{a_i} {da \over a}\exp\left[{3\over2}
\int^{a}_{a_i}{\gamma(a')\over a'}da'\right]~~~.
\label{aa12}
\ene
{}From $\ga$, immediately follows the value of the e-fold number
$N=\log(a_{f}/a_{i})$.
Finally, we notice that this reconstruction procedure can also be reversed
in order to obtain the properties of the cosmological fluid (equation of state)
once that the inflationary potential, or the dynamics of the scalar fields
are known. We will show in the following one example of this inverse treatment.

To end this section it is worth pointing out the connection existing between
the slow-roll approximation and this description given in terms of $\ga$. As it
is well-known, the slow-roll regime can be conveniently characterized in terms
of two parameters $\epsilon$ and $\eta$, defined in \cite{4a}, by the
conditions
$\epsilon$, $|\eta|<<1$. These two quantities are
simply connected to deviation from the flat Harrison-Zeldovich scalar
perturbation spectrum, and according to the standard approach, this connection
allows to reconstruct perturbatively the inflaton potential. By definition
the expressions of $\epsilon$ and $\eta$ in terms of $\ga$  result to be
\begin{equation}
\epsilon = {3 \over 2} \left( { p \over \rho} +1\right)   = { 3 \over 2}
\ga~~~,
\label{c24}
\end{equation}
\begin{equation}
\eta  = {3 \over 2} \left( {\partial p \over \partial \rho} +1\right)   =
{ 3 \over 2} \ga - { 1 \over 2 } { d \log(\ga) \over d \ln}~~~.
\label{d24}
\end{equation}
Interestingly, from (\ref{c24}) and (\ref{d24}) $\epsilon$
represents the adiabatic index, whereas $\eta$ is linearly related to
the squared value of the generalized sound speed.

\section{Density perturbations and relevant examples of equations of state.}

A typical feature of inflationary models is to provide a general scheme for the
production of primordial fluctuations. This aspect is of particular importance,
since the experimental information on the CMBR temperature fluctuations,
and large scale galactic structure can constraint, to some extent, the
form of the inflationary potential. In this section we will stress how,
starting from the adiabatic index $\ga$, perturbation spectra are easily
reconstructed. As in \cite{4a} we introduce the perturbation amplitudes $A_S$
and $A_G$ corresponding, respectively, to quantum fluctuation of the inflaton
field (scalar perturbations) and metric (tensorial perturbations)
\beqa
A_S & = & \sqrt{2 \over \pi} G ~H^2(\phi) \left|
{\delta H(\phi) \over \delta \phi} \right|^{-1}~~~,\label{d25a}\\
A_G & = & \sqrt{\frac{G}{2 \pi^2}}~ H(\phi)~~~,
\label{d25}
\enqa
where $H(\phi)= \sqrt{8 \pi G \rho/3}$ is the Hubble parameter.

In particular, from (\ref{c24}) the ratio $A_G/A_S$ is related to $\ga$
in a simple way
\bee
\frac{A_G}{A_S} = \sqrt{\epsilon} = \sqrt{{ 3 \over 2}~\ga}~~~,
\label{d27}
\ene
and therefore it is small in the slow-roll regime.
In the same limit the scale dependence of the spectra can be expressed in
terms of $\ga$
\beqa
1-n_{S} \equiv
{d \log\left[A_S^2(\lambda)\right] \over d \log(\lambda/\lambda_{0})}
& = & 4 \epsilon(a) - 2 \eta(a)
= 3 \ga +  { d \log(\ga) \over d \ln}~~~,\label{d28a} \\
n_G \equiv {d \log\left[A_G^2(\lambda)\right] \over d
\log(\lambda/\lambda_{0})} & = & 2 \epsilon(a) = 3 \ga~~~,
\label{d28}
\enqa
where the parameters $\epsilon(a)$, $\eta(a)$ have to be evaluated at the value
for $a$ when the scale $\lambda$ goes outside the Hubble radius during the
inflationary era.  To this end we recall that the physical scale
$\lambda(\phi)$, decoupled when the value of the scalar field was $\phi$, grew
between the time of horizon crossing and today by a factor \cite{4a}
\bee
\lambda(\phi) = H^{-1}(\phi) \frac{a_0}{ a(\phi)}~~~,
\label{d28bis}
\ene
where $H(\phi)$ and $a(\phi)$ represent, respectively, the value of the
Hubble radius and of the scale factor at decoupling, and $a_0$ is the
present value of the scale factor. Since
\bee
a(\phi)= a_f ~\exp[ - N(\phi)]~~~,
\label{d29}
\ene
with $N(\phi)$ the number of e-fold between the value $\phi$ and the
the end of inflation, from (\ref{d28bis}) and (\ref{d29}) we get
\bee
\lambda(\phi)= \frac{ \exp[N(\phi)]}{H(\phi)} \frac{a_0}{a_f}~~~.
\label{d31}
\ene
The function $N(\phi)$ can be obtained using the definition of the Hubble
parameter and the equation of motion. Differentiating in fact $H^2$ with
respect
to $\phi$
\bee
\frac{\delta H^2 }{\delta \phi} =
2 \dot{H} H \dot{\phi}^{-1} = \frac{8 \pi G}{3}
\left( \ddot{\phi} + \frac{\delta V(\phi)}{\delta \phi} \right)~~~,
\label{new1}
\ene
which, by virtue of (\ref{a3}), gives
\bee
\frac{\delta H}{\delta \phi}  = - 4 \pi G \dot{\phi}~~~.
\label{new2}
\ene
Therefore, if $t_f$ is the value of time at the end of the inflationary phase,
using (\ref{new2})
\bee
N(\phi) = \int_{t(\phi)}^{t_f} H(t') dt' =
\int_{\phi}^{\phi_f} { H(\phi') \over \dot{\phi'}} d \phi' =
- 4 \pi G \int_{\phi}^{\phi_f} H(\phi') \left({\delta H(\phi')
\over \delta \phi'}\right)^{-1} d \phi'~~~,
\label{d30}
\ene
If we now differentiate (\ref{d31}) with respect to $a$ we
find, using (\ref{a12}) and (\ref{d30})
\beqa
\frac{d log( \lambda / \lambda_0 )}{d \ln} & = & \left[ 4 \pi G
 H \left( \frac{\delta H}{\delta \phi} \right)^{-1} -
H^{-1} \frac{\delta H}{\delta \phi}
\right] \frac {d \phi}{d\ln} \nonumber \\
& = & \left[ 4 \pi G
 H \left( \frac{\delta H}{\delta \phi} \right)^{-1} -   H^{-1}
\frac{\delta H}{\delta \phi}
\right] \sqrt{ \frac{3 \gamma(a) }{8 \pi G} } {\mbox{sign}} \left( \frac{d
\phi}{da} \right) ~~~,
\label{new3}
\enqa
which, from the definition of $A_S$ and $A_G$ in (\ref{d25a}) and (\ref{d25}),
can also be cast in the form
\bee
\frac{d \log(\lambda/\lambda_{0})}{d \ln} =
{\mbox{sign}}\left({ \delta H(\phi) \over \delta \phi}\right)~
{\mbox{sign}} \left( \frac{d \phi}{da} \right) ~
\left[ \frac {A_S}{A_G} -
\frac{A_G}{A_S} \right] \sqrt{ \frac{3 \gamma(a)}{2} } ~~~.
\label{new4}
\ene
Finally, from (\ref{d27}) we get the simple result
\bee
\frac{d \log(\lambda/\lambda_{0})}{d \ln} = - {\mbox{sign}}\left({ d
H(a) \over d a}\right) \left( \frac{3}{2} \gamma(a) -1 \right) =
\frac{3}{2} \gamma(a) -1~~~,
\label{d32}
\ene
since combining (\ref{a-1}) and (\ref{a0}), for $\gamma(a)>0$
\bee
\dot{H} = - 4 \pi G \gamma \rho < 0~~~,
\label{new5}
\ene
and during the expansion $\dot{a}>0$, implying that $dH/da<0$.\\
Equation (\ref{d32}) allows us to express directly the perturbation spectra
as functions of $a$: using (\ref{d28a}) and (\ref{d28})
\beqa
{d \log\left[A_S^2(a)\right] \over d \ln}  & = &
\left[ 3 \ga + { d \log(\ga) \over d \ln} \right]~
\left( {3 \over 2} \ga -1 \right) ~~~,
\label{d33a} \\
{d \log\left[A_G^2(a)\right] \over d \ln}  & = & 3 \ga~
\left( {3 \over 2} \ga -1 \right) ~~~.
\label{d33}
\enqa
To end this section with two relevant examples, we will deduce the expressions
for the adiabatic index $\ga$, for the most simple "chaotic"
potential, $V(\phi) = \mu^2 \phi^2$, and for the potential leading to the flat
Harrison-Zeldovich spectrum for scalar perturbations.

In the slow-roll limit the case $V(\phi) = \mu^2 \phi^2$ may be easily
solved, considering the evolution of the scalar field once it has reached
{\it limit velocity} condition, and neglecting the kinetic energy contribution
to the Hubble parameter. From (\ref{a-1}) and (\ref{a3}) one obtains
\beqa
\phi(t) & = & \phi_0 - \frac{\mu}{\sqrt{6 \pi G}} t ~~~,\\
a(t) & = & a_0 ~
\exp \left\{ \frac{ \sqrt{8 \pi G} \mu }{\sqrt{3}} \left[ \phi_0 t
- \frac{\mu}{ 2 \sqrt{6 \pi G } } t^2 \right] \right\}~~~.
\label{d35}
\enqa
The end-point of the inflationary phase is determined by the condition
$\ddot{a}(t) = 0$, giving the final value of the inflaton field
$\phi_f$ and the value of $N$
\beqa
\phi_f & = & (4 \pi G)^{-1/2}~~~, \label{d37a}\\
N & =&  2 \pi G \phi_0^2 - {1 \over 2}~~~.
\label{d37}
\enqa
Using the previous relations we obtain the adiabatic index
\bee
\gamma(a)=
{ 2 \over 3} \left[ 2N+1 - 2 \log\left({ a \over a_{i}} \right)
\right]^{-1}~~~,
\label{d38}
\ene
which shows that the inflaton field starts its evolution with condition
which differs from {\it perfect slow-roll} for term $1/N$ ($\gamma(a_{i})
\approx 1/3N$), increases its velocity and finally exit from the
inflationary phase $\gamma(a_{f})=2/3$).\\
Starting from (\ref{d38}) we can now easily obtain the perturbation
amplitudes. We first notice that in the present approximation $\eta=0$, so
$1-n_{S} = 2~n_{G}$. From this by using (\ref{d32})-(\ref{d33}) we obtain
at the lower order the well-known expressions \cite{4a}
\begin{eqnarray}
A_{S}(\lambda) & = & A_{S}^{0} \left[ 1 + { 2 \over 2N + 1} \log\left(
{\lambda \over \lambda_{0}}\right) \right]~~~,\label{d38a}\\
A_{G}(\lambda) & = & A_{G}^{0} \left[ 1 + { 2 \over 2N + 1} \log\left(
{\lambda \over \lambda_{0}}\right) \right]^{1/2}~~~,\label{d38b}
\end{eqnarray}
where, from relations (\ref{d25a}) and (\ref{d27}), since $\lambda_0$ is
defined
as the scale which decouples at $\phi_0$
\beqa
A_S^0 & = &  \frac{ 4 G^{3/2} \mu \phi_0^2}{ \sqrt{3} }~~~, \label{new6} \\
A_{G}^0 & = &  \frac{A_{S}^0}{\sqrt{2N+1}}~~~. \label{new7}
\enqa
To reconstruct the Harrison-Zeldovich potential we remind that in order to have
a scale independent fluctuation amplitude, it is necessary that $2 \epsilon(a)
= \eta(a)$, or equivalently that the r.h.s. of (\ref{d28a}) identically
vanishes. Hence, from this condition we straightforwardly get the differential
equation for $\ga$
\bee
{ d \ga \over d \ln} = -3 ~\gamma^2(a)~~~,
\label{d39}
\ene
whose solution is
\bee
\gamma(a) = { 2 \over 3}
\left[ { 2 \over 3 \gamma_{0}} + 2 \log\left( { a \over a_{i}}
\right) \right]^{-1}~~~,
\label{d40}
\ene
with ${\gamma}_0$ the initial value of $\ga$. In particular if
${\gamma}_0=0$ the adiabatic index remains constant and it is easily
seen from (\ref{a15}) that the potential reduces to a constant, and
the amplitude of scalar perturbations tends to infinity. Notice that
$\ga$ remains always under the inflationary threshold of $2/3$, leading,
in absence of other fields to eternal inflation.

Evaluating the expression of $V(a)$ and $\phi(a)$ we find for
${\gamma}_0 \neq 0$:
\bee
V(\phi) = \frac{\rho_0 }{3(1 - \hat{\gamma}_0)} \left\{ 3
\frac{\bar{\phi}^2}{\phi^2} -
\frac{\bar{\phi}^4}{\phi^4} \right\}~~~,
\label{d41}
\ene
with $\bar{\phi} = (4 \pi G)^{-1/2}$. Eq. (\ref{d41}), obtained by suitably
choosing $\phi(0)$, recovers the well-known form of the only potential leading
to the Harrison-Zeldovich spectrum for scalar perturbations \cite{4a}.

\section{$1/$e-fold number expansion}

As well-known an inflationary phase in the universe evolution is able to
solve a number of important problems occurring in the standard cosmology
provided that the final e-fold number, $N=\log(a_f/a_i)$, is large enough. A
reasonable estimate for this quantity would be $N \approx 100$. The value of
$N$
is completely fixed once the equations of motion and the initial conditions
are assigned. By using the approach described in the previous sections,
we will adopt here a different point of view,
namely we will formally look at $N$ as a quantity in terms of which
the potential and the field dynamics can be parameterized.
Since, as already mentioned, its value is very large it is quite natural
to expand in $1/N$ all physical quantities, building in this way a
perturbative approach to inflaton dynamics. Notice that $N$ is a
{\it global} quantity, being related to the total growth of the scale
factor during the inflationary phase, and the $1/N$ expansion is
completely independent on whether slow-roll conditions are satisfied or
not: this is quite different from the usual perturbative approach
in terms of the slow-roll parameters $\epsilon$ and $\eta$ \cite{4a},
which allows for an accurate description only in the
slow-roll regime.

In order
to simplify our analysis let us introduce the dimensionless quantities $y$ and
$\gy$ defined by
\bee
y\equiv{1\su N}\log\left({a\su a_i}\right)~~~,
\label{a13}
\ene
and
\bee
\gy\equiv 1-{3\su2}\g(a(y))~~~.
\label{a14}
\ene
By definition, the $y$ variable during the inflationary epoch is bound to be $0
\le y\le 1$, and the starting and ending points of inflation correspond to
$y=0$ and $y=1$ respectively. Actually, in this interval $0\leq \gy \leq
1$. The definition of the variable $y$ is quite natural by observing that
it would simply correspond to $H t$
for a pure exponential expansion. Eqs. (\ref{a9}) and (\ref{a12}), when
expressed in terms of $y$, suggest to introduce two dimensionless quantities,
obtained by properly rescaling potential and field configuration
\bee
\widehat{V}(y) \equiv \left[{V(y) \over \rho_{0}}\right]^{1/N}
= \left[ { 1 \over 3} (2 + \gy )\right]^{1/N}\exp \left\{
-2 \int_{0}^{y} [ 1- \gyp]~dy' \right\}~~~,
\label{a15}
\ene
and
\bee
\hat{\f}(y) \equiv \sqrt{ 4 \pi G} \left[ { \f(y) - \f(0) \over  N}
\right]= \pm \int_{0}^{y} \sqrt{1- \gyp}~dy'~~~.
\label{a16}
\ene
We first note that $\widehat{V}(y)$ results to be an analytical
function of $1/N$ provided that $\gy$ is analytical as well:
in particular, expanding (\ref{a15}) in $1/N$ up to the first order
\bee
\widehat{V}(y) \simeq  \exp \left\{-2 \int_{0}^{y} [ 1- \gyp]~dy'
\right\}\left[1+{1 \over N} \log\left( { 2 + \gy\over 3}\right) \right]~~~,
\label{a17}
\ene
$\widehat{V}(y)$ gets at zero-th order a very simple functional dependence on
$\gy$. In terms of the quantities (\ref{a14}), (\ref{a15})
and (\ref{a16}) we can also rewrite the equation of motion (\ref{a3}) in
the following form
\bee
{1 \over N} \hat{\f}''(y) + [ 2 + \gy] \hat{\f}'(y) +
{ 1 \over 2} [2 + \gy] { 1 \over \widehat{V}(y)} { \delta \widehat{V}(y)
\over \delta \hat{\f}}=0~~~,
\label{b17}
\ene
where prime denotes differentiation with respect to $y$. In the large $N$ limit
it reduces simply to
\bee
\hat{\f}'(y) +
{ 1 \over 2} { 1 \over \widehat{V}(y)} { \delta \widehat{V}(y)
\over \delta \hat{\f}}=0~~~,
\label{c17}
\ene
which corresponds to the zero-th term of the $1/N$ expansion
for the $\hat{\f}$-field equation of motion; in particular
note that, consistently with the expansion, (\ref{a16}) and (\ref{a17})
at the lowest order in $1/N$, exactly solve (\ref{c17}). The first order
differential equation (\ref{c17}) provides an {\it improved}
slow-roll approximation for the scalar field dynamics. It is in fact
the analogous of the slow-roll limit for
the classical equation of motion for $\f$, but it does also contain
a contribution coming from the second-time derivative of the field.
Had we neglected the second time derivative term, as in the usual slow-roll
approximation, we would have rather obtained
\bee
\hat{\f}'(y) +
{ 2 + \gy \over 6} { 1 \over \widehat{V}(y)} { \delta \widehat{V}(y)
\over \delta \hat{\f}}=0~~~,
\label{b17a}
\ene
which reduces to (\ref{c17}) only in the case of a perfectly flat
potential ($\gy=1$). Consequently, we point out that slow-roll
approximation implies large $N$, but, viceversa, there are efficient
inflationary models, for which $1/N$ expansion is still reliable, which
do not satisfy slow-roll conditions. This conclusion can be qualitatively
got from the expression of the e-fold number $N$: we first remind that,
using (\ref{new2}), $\epsilon$ and $\eta$ can also be written in the
form\footnote{This is actually the form in which they were introduced in
reference \cite{4a}.}
\beqa
\epsilon & = & \frac{1}{4 \pi G} \left( \frac{1}{H} \frac{ \delta H}{ \delta
\phi} \right)^2 \label{25} \\
\eta & = & \frac{1}{4 \pi G} \frac{1}{H} \frac{ \delta^2 H}{ \delta
\phi^2} \label{26}
\enqa
so from (\ref{d30})
\bee
N = - 4 \pi G \int_{\phi_0}^{\phi_f} \frac{ d \phi}{\sqrt{\epsilon(\phi)}}
\label{27}
\ene
If $\epsilon$ is small one expects the integral of $(\sqrt{\epsilon})^{-1}$,
as a function of $\phi$, to be large, though this is only a sufficient
condition:
it is in fact conceivable to still have $N>>1$ even if $\epsilon$ is close to
1 somewhere in the interval $[min\{\phi_0,\phi_f\},max \{\phi_0,\phi_f\}]$.\\
The relationship between the slow-roll and the $1/N$ expansions can be however
formalized by introducing a new parameter
\bee
\nu = \frac{ \hat{\f}''}{N (2 + \gy) \hat{\f}'}
\label{28}
\ene
The condition $\nu<<1$ corresponds to the possibility of neglecting the $1/N$
term in the dimensionless equation of motion (\ref{b17}), which therefore
reduces to (\ref{c17}): we will hereafter refer to the latter as the $1/N$
approximation and to the expansion of the dynamical quantities in terms of the
two parameters $1/N$ and $\nu$ as the $1/N$ expansion. Using (\ref{a16}),
(\ref{c24}) and (\ref{d24}) one easily gets
\bee
\nu = \frac{\epsilon - \eta}{3 - \epsilon}
\label{29}
\ene
Remarkably $\nu$ contains all power terms in $\epsilon$. This relation shows
that if $\epsilon$ and $\eta$ are small, namely if the slow-roll approximation
holds, then also $\nu$ is small and the $1/N$ approximation is legitimate;
viceversa, if $\nu$ is small, this only implies that the difference $\epsilon -
\eta$ is small, while both $\epsilon$ and $\eta$ can be quite large: in this
case the $1/N$ approximation is still valid, while the slow-roll one breaks
down. Summarizing, $\mbox{slow-roll} \Rightarrow 1/N$, but
$1/N \not\Rightarrow \mbox{slow-roll}$, so, as we wanted to show, the $1/N$
expansion seems to be more widely
applicable to inflationary dynamics since it grasps its main feature (large
e-fold).\\
It is interesting at this point to look for the potential satisfying the
condition $\nu=0$, i.e. for which the $1/N$ approximation already gives an
exact solution: from (\ref{c24}) and (\ref{d24}) it follows that $\gamma$
is a constant, $\gamma(a)=\gamma_0= 2 \epsilon_0/3 > 0$, so, using (\ref{a15})
and (\ref{a16}), it is easy to obtain the expression of $V(\phi)$
\bee
V(\phi) = \rho_0 \left( 1 - \frac{\epsilon_0}{3} \right)
\exp \left[ - \sqrt{4 \pi G \epsilon_0} ~| \phi - \phi_0 | \right]
\label{30}
\ene
Let us now finally consider the slow-roll regime in further details.
It is well-known that the consistency of this
approximation for the scalar field dynamics requires \cite{9a}
\begin{eqnarray}
\left| { 1 \over V(\phi)} { \delta V(\phi) \over \delta \phi} \right|
& << &\sqrt{48 \pi G} ~~~,
\label{a22}\\
\left| { 1 \over V(\phi)} { \delta^2 V(\phi) \over \delta \phi^2} \right|
& << &24 \pi G~~~.
\label{a23}
\end{eqnarray}
Also in this case the rescaled quantities (\ref{a14}), (\ref{a15})
and (\ref{a16}) allow to rewrite (\ref{a22}) and (\ref{a23}) in a
more suitable form
\begin{equation}
\left| { 1 \over \widehat{V}(
\hat{\phi})} { \delta \widehat{V}(\hat{\phi}) \over \delta
\hat{\phi}} \right| << 2 \sqrt{3}~~~,
\label{b22}
\end{equation}
and
\begin{equation}
\left| { 1 \over \widehat{V}^2(\hat{\phi})}
{\left({ \delta \widehat{V}(\hat{\phi}) \over \delta \hat{\phi}} \right)}^2
+ { 1 \over N} { 1 \over \widehat{V}^2(\hat{\phi})}
\left[ \widehat{V}(\hat{\phi})
{ \delta^2 \widehat{V}(\hat{\phi}) \over \delta \hat{\phi}^2} -
{\left({ \delta \widehat{V}(\hat{\phi}) \over \delta \hat{\phi}} \right)}^2
\right] \right| << 6~~~.
\label{b23}
\end{equation}
Remarkably, the transformation introduced in (\ref{a13})-(\ref{a16}),
besides allowing us to perform the $1/N$ expansion of the physical quantities,
has the nice property to leave (\ref{a22}) formally unchanged.
Furthermore, we notice that at the zero-th order in $1/N$
expansion, (\ref{b22}) and (\ref{b23}) are essentially equivalent and are
both satisfied provided that
\begin{equation}
\left| { 1 \over \widehat{V}(
\hat{\phi})} { \delta \widehat{V}(\hat{\phi}) \over \delta
\hat{\phi}} \right|  <<  \sqrt{6}~~~.
\label{b24}
\end{equation}

\section{Conclusions}

Following the reconstruction procedure of the inflaton potential
based on the knowledge of the adiabatic index for the cosmological fluid
as a function of the universe scale factor $a$ \cite{6a},
we have studied the slow-roll regime and in particular the connection existing
between $\ga$ and the perturbation parameters $\epsilon$ and $\eta$ \cite{4a}.
In this framework we have also computed the scalar and tensorial perturbation
spectra as a function of $\ga$, and as an example we have analyzed two simple
cases, namely the most elementary chaotic model and the Harrison
Zeldovich potential; for both this cases the adiabatic index $\gamma(a)$
is found to be a simple rational function of $\log(a/a_{i})$.
Interestingly, the simple connection occurring between
the adiabatic index and the slow-roll parameters $\epsilon$ and $\eta$
provides once that scalar and tensorial perturbation amplitudes are known,
to completely characterize the cosmological fluid.
Finally, by applying the reconstruction procedure
we have developed a formalism in which it is natural
to perform a $1/N$ expansion ($N$ is the e-fold number) of the most
relevant physical quantities characterizing the inflationary dynamics.
This expansion seems to be more general than the one given in terms of the
slow-roll parameters, being applicable also in cases when even though
the e-fold number is large, slow-roll conditions are not satisfied.
In particular this approach is based on the idea of using two new
parameters, $1/N$ and $\nu$, in describing inflationary dynamics, instead of
the slow-roll ones $\epsilon$ and $\eta$.
At zero-th order in this expansion an improved slow-roll approximation is
straightforwardly obtained.

\end{document}